# *Wave-based quasi-digital logic operations*

*Ross Glyn MacDonald[1], Alex Yakovlev[2] and Victor Pacheco-Peña[1*]*

[1]*School of Mathematics, Statistics and Physics, Newcastle University, Newcastle Upon Tyne, NE1 7RU, United Kingdom*
[2]*School of Engineering, Newcastle University, Newcastle Upon Tyne, NE1 7RU, United Kingdom*
*\*email: victor.pacheco-pena@newcastle.ac.uk*

**Electromagnetic wave-based computing has emerged as an exciting paradigm with the potential to enable high-speed, parallel operations. In conventional computing, elementary logic gates, such as $AND$, $OR$, $NOT$ and $XOR$, form the building blocks of larger interconnected logic circuits. These operations are inherently non-linear processes, which may be challenging to implement using an electromagnetic wave-matter based systems. In this work, we discuss how one may instead *emulate* the functionality of certain logic gates by using quasi-digital linear systems (in this case networks of interconnected parallel plate waveguides). This is done by carefully designing the encoding scheme of the input bits (how the input state of each bit maps to an incident electromagnetic signal) and the high/low classification regions of the output signal. To demonstrate this approach, full-wave numerical simulation of elementary 2-input operations are presented. This technique is then extended to many-to-one and many-to-many operations such as an *N*-input $AND$, half, full and a 2-bit full-adder. We envision that this quasi-digital linear computing technique may help enable new opportunities for electromagnetic wave-based computing.**

# Introduction

Semiconductor-based computing technologies are the backbone of the digital world. One component of this success is the exceptional scalability[1] of metal-oxide-semiconductor field-effect transistors (MOSFETs), which has enabled the production of increasingly dense processors as semiconductor manufacturing technology has advanced over time[2–4]. However, as device sizes approach atomic limits, increasing manufacturing costs and the impact of quantum tunneling make it challenging to maintain this rapid rate of advancement[5,6]. While research and innovation to overcome these challenges is ongoing, there has also been an interest in exploring further computing paradigms[7–11]. Examples include computing with solitons[12–16], quantum computing[8,17–19], nature inspired computing[20–22], advanced packaging[23,24] and wave-matter-based computing[25–32], to name a few.

In recent years, it has been shown how electromagnetic (EM) wave-matter interactions may have the potential to overcome some of the current challenges by harnessing the advantages of light for computing purposes[15,25,28,33,34]. Interesting features of wave-matter-based computing structures include their inherent high speed[28,35] (transport of information at the speed of light inside the medium) and their potential for parallelization associated with orthogonal modes or different operating frequencies, among other examples[28,36–39]. Another feature of EM wave-matter computing is that the need for EM-electronic conversions may be reduced, as all the calculations can be done fully via wave-matter interactions[40–42]. In the realm of analogue computing[25], EM processors capable of performing fundamental operations such as switching and routing[35,43–47] of information or key mathematical operations including differentiation[48–52], integration[29,53], vector matrix multiplication[54–56] and convolution[57–60], have been reported. Additionally, analogue processors have also been designed to perform comparably intensive task such as ordinary differential[60–62] or partial differential equation solving[63–65].

Digital computing has also benefited from the application of EM waves[27,28]. In this realm, various structures that can emulate the performance of digital processes, such as decision trees[35,43,44] or logic gates[27,28], using a fully analogue EM wave-matter-based system have been proposed. These studies can be classified into two categories: those that exploit non-linear processes[66–76] and those that leverage linear processes[77–86]. In general, digital computing requires a degree of non-linearity[87,88] to enable a mapping between input and output values and, while there are known limits on the operations that can be performed by a purely linear system[77,78,87], designing large interconnected non-linear systems also poses its own challenges[28,89]. Interestingly, it has been shown that certain digital operations, such as elementary logic gates[77,78,80,90] (*NOT*, *AND*, *OR*, *XOR*, *NAND*, *NOR*, and *XNOR*), can be emulated in linear systems by carefully tailoring the superposition of input signals[77,78,80,90,91]. However, without non-linearity the output signals may have a wide range of potential amplitudes and phases, which also poses a challenge when cascading logic operations[92].

In this work, a comprehensive theoretical and numerical investigation of digital logic operations implemented using linear systems is presented. These operations are referred to as linear quasi-digital logic gates. We adopt this term to note that our aim is to *emulate* logic operations while emphasizing that our systems are linear. It is shown how, by optimizing both the encoding scheme of the input signals (how high and low input states map to the wave characteristic of the input signals) and the classification domains (threshold values which classify output signals as representing a high or a low output state) of the output signals, the performance of many-input and non-elementary quasi-logic gates can be emulated. To illustrate this, the operation of an EM wave-matter-based *N*-input *AND* gate (where *N* is the number of input bits), half-adder, full-adder and two-bit full-adder is discussed. Numerical simulations of these operations, implemented using interconnected networks of parallel plate waveguides, are presented throughout. In these examples, the input encoding schemes, and the dimensions of the waveguide structures, are optimized to maximize the contrast ratio of the operation (separation between high and low classification domains), thereby reducing the likelihood of misclassification. We envision that this work may aid in the development of EM wave-matter-based logic gates as the principles discussed throughout may be transferred to different frequencies and waveguide-based structures.

## Results

**Quasi-digital linear logic gates: Operating principle**

The premise of an EM wave-based linear quasi-digital logic gate is to harness the constructive and destructive interference of multiple incident EM waves to perform computing operations[77,78]. In this realm, each incident signal can be representative of a single bit of data, with the high/low state of this bit encoded into certain wave characteristics of the EM signal. For instance, information may be encoded into the magnitude or phase of an input signal with a low or high input state represented by an input magnitude of 0 or 1 (arbitrary units) or by an input phase of 0 or $\pi$ radians, respectively. Information could also be encoded into the wavelength of an input signal[93,94] (outside the scope of this manuscript). Here, the types of operations that can be obtained by exploiting the linear superposition of the scattered signals produced by multiple monochromatic incident waves are explored (all at the same frequency). In this context, a hypothetical $N$-input operation, with $a = [1,2, ..., N]$ as the labels of the input bits, will require $N$ input signals $x_a$, each excited at a unique point in space. For example, $x_1$ and $x_2$ (signals representative of bit 1 and bit 2, respectively) could be excited at two different waveguide ports within a network of waveguides. To encode the input bits, $x_a$ is allowed to take one of two values: $H_a$ or $L_a$ when bit $a$ is high or low, respectively. Then, for a linear system, the signal seen at a certain output $y_{out}$ (such as at a port of a waveguide network) is the superposition of the scattered output signals produced by each of the input signals, as follows:

$$y_{out} = A_{out,C} x_C + \sum_a^N A_{out,a} x_a \qquad (1)$$

where $A_{out,a}$ is the transmission coefficient between where $x_a$ is excited and where $y_{out}$ is observed. Additionally, note that there is another term outside the summation. This term, $A_{out,C} x_C$, represents a scattered signal (associated to a scattering parameter $A_{out,C}$) produced by a monochromatic incident signal $x_C$ (here used as a control signal) which can be applied independently of the state of the input bits.

Now, how can one exploit Eq. 1 to emulate logic operations? As it is known, for an operation to be successfully realised, the signals representing high and low output states must be clearly distinguishable from one another. Therefore, Eq. 1 can be used to emulate a quasi-digital logic operation within a linear structure if it allows for distinct (not overlapping) high and low classification regions (with associated threshold values) to be defined. Now, there are three parameters which may be used to control the outcome of Eq. 1. The first is the encoding scheme of the input bits (chosen values of $L_a$ and $H_a$). The second is the structure used to control the superposition of the signals ($A_{out,a}$ and $A_{out,C}$ values). The last is the chosen value of control signal $x_C$ which will influence all input combinations equally. A schematic example of the concept of a hypothetical quasi-digital linear logic gate exploiting these three control mechanisms is presented in Fig. 1.

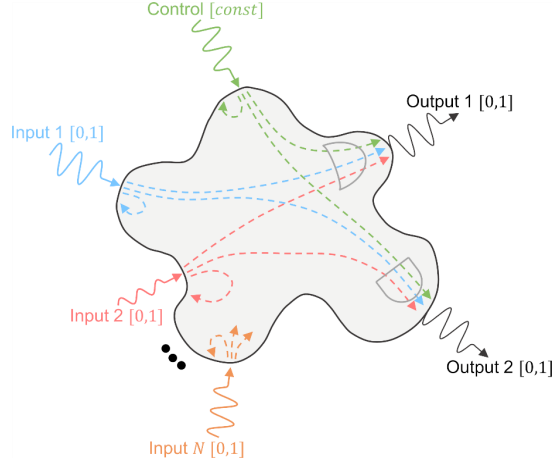

**Fig 1: Schematic representation of the operating principle of quasi-digital linear logic gates.** A hypothetical quasi-digital linear logic gate implementing a many-input OR and many-input $AND$ gate simultaneously by controlling: 1) encoding scheme of $x_a$, 2) The structure of the linear logic gate and 3) the applied control signal $x_C$.

To evaluate the performance of a quasi-digital linear logic gate, the contrast ratio $C_R$ of the operation can be defined as follows[27,83,95]:

$$C_R = 20 \log_{10}\left(\frac{Th_H}{Th_L}\right) \quad (2)$$

where $Th_H$ and $Th_L$ are the threshold values of $y_{out}$ for the high and low classification regions, respectively. $C_R$ is the separation between the two classification regions in dB. It is also feasible to design a quasi-digital operation in which $Th_L > Th_H$, as will be shown later in this manuscript. Due to this, in this manuscript, we will consider the magnitude of $C_R$ as we are interested in its absolute value. Moreover, based on Eq. 2, if $Th_H$ and $Th_L$ cannot be defined for a set of input parameters and a desired operation (i.e. there is an overlap between the high and low classification regions), then the operation cannot be resolved[78]. In this case, the high or low state of some output signals (those which are within the overlapping region) would be ambiguous, as expected. Due to this, the design of a quasi-digital linear logic gate would prioritize a maximization of $C_R$ so that the potential misclassification of output signals can be minimized[77].

To begin with, let us first consider one possible implementation of some elementary 2-input quasi-logic operations: $XOR$, $OR$, $AND$ and $NAND$. Numerical simulations of these operations are presented in Fig. 2. For each operation, the structure, encoding scheme and control signal has been designed to maximize $C_R$. For completeness an example implementation of the remaining elementary logic gates ($NOT$, $NOR$ and $XNOR$) can be found in the supplementary materials. To realise these operations, the splitting and superposition of signals at a series junction between parallel plate waveguides is exploited. The required input signals ($x_1$ and $x_2$) are excited by waveguide ports placed at the outer ends of the left and right waveguides, respectively. The fundamental TEM mode is used throughout this work. The $XOR$ and $OR$ operations (Fig. 2b left and right panels, respectively) are implemented without a control signal (structure from Fig. 2a). However, for the $AND$ and $NAND$ operations (Fig. 2d left and right panels, respectively) the control signal is introduced by a waveguide port exciting the top waveguide (see structure from Fig. 2c). The output signal, $y_{out}$, is extracted from the bottom waveguide using a waveguide port.

All waveguides in both structures from Fig. 2 have the same characteristic impedance $Z_0$ and are filled with air ($\varepsilon_r = \mu_r = 1$). Considering the splitting and superposition of waves at the junction and accounting for the phase change due to propagation through the waveguides, $y_{out}$ is calculated as[94]

$$y_{out} = \gamma_3 e^{i\beta l_1} x_1 + \gamma_3 e^{i\beta l_2} x_2 \tag{3}$$

for the structure presented in Fig. 2a. Here, $l_1 = d_1 + d_{out}$ and $l_2 = d_2 + d_{out}$ are the total waveguide lengths from inputs 1 and 2 to the output. The parameters $d_1$, $d_2$ and $d_{out}$ represent the length of the waveguides connecting ports 1, 2 and the output to the centre of the junction, respectively. Note that Eq. 3 (along with Eqs. 4, 8-11, which describe the output signals) is derived using transmission line (TL) theory[96]. Hence, $l_1$ and $l_2$ are the lengths of the TLs modelling the waveguides. These values do not account for sharp waveguide bends (as those present in the numerically simulated structures) which may introduce minor variations in the simulated path lengths. Nevertheless, this discrepancy is negligible, as it is demonstrated by the agreement between the theoretical and numerical results presented throughout this work. Finally, $\gamma_3 = 2/3$ is the transmission coefficient of a 3-waveguide junction with equal impedances[35] and $\beta = 2\pi/\lambda_0$ ($\lambda_0 = 30$ mm) is the propagation constant. Following the same approach, for the structure in Fig. 2c, $y_{out}$ is instead:

$$y_{out} = \gamma_4 e^{i\beta l_1} x_1 + \gamma_4 e^{i\beta l_2} x_2 + \gamma_4 e^{i\beta l_c} x_c \tag{4}$$

where $l_c = d_c + d_{out}$ is the total waveguide length from the control input to the output port ($d_c$ is the length of the waveguide between the control input and the junction) and $\gamma_4 = 1/2$ is the transmission coefficient of a 4-waveguide junction with equal impedances[35].

**Elementary linear logic gates: Numerical Results**

To implement an $XOR$ operation using Eq. 3, the following encoding scheme is chosen: $L_1 = L_2 = 1 \angle 0$ rad and $H_1 = H_2 = 1 \angle \pi$ rad (normalized with respect to the $L_1$ input signal) with the following structural parameters: $d_1 = 3\lambda_0/2$, $d_2 = \lambda_0$. These values are chosen as, when using Eq. 3, the signals representing bit 1 and bit 2 ($x_1$ and $x_2$, respectively) will destructively/constructively interfere when the two bits are in the same/opposite input states, respectively, thus emulating an $XOR$-like operation. Full-wave numerical simulation results of this scenario calculated using the commercial simulation software CST Studio Suite® are presented in Fig. 2b (left panels). The power distribution inside the waveguide network and the output signals for all four combinations of the input signals representing bits 1 and 2 are presented in the top four and bottom panels respectively. i.e. $[0,0] \rightarrow [L_1, L_2]$, $[0,1] \rightarrow [L_1, H_2]$, $[1,0] \rightarrow [H_1, L_2]$, $[1,1] \rightarrow [H_1, H_2]$, where the values in the left brackets represent the input state of bit 1 and bit 2, respectively, and the values in the right brackets are the corresponding values of $x_1$ and $x_2$, respectively. The numerically calculated $|y_{out}|$ values were: $|y_{00}| = |y_{11}| = 0.00155$ and $|y_{01}| = |y_{10}| = 1.332$ which is in agreement with the theoretical values of $|y_{00}| = |y_{11}| = 0$ and $|y_{01}| = |y_{10}| = 4/3$ from Eq. 3. Using Eq. 2, the contrast ratio $C_R$ of this operation is 58.69 dB with $Th_H = 1.332$ and $Th_L = 0.00155$, respectively. However, in a real system this is expected to be lower as threshold values would need to be defined including tolerances for signal noise[77,97]. Interestingly, while this structure indeed emulates an $XOR$ operation, it differs from an ideal logic gate in that the classification of output states is different than the encoding of input states. In this case, encoding into the phase of the input signals and classifying based on the magnitude of the output signals. This is another reason why we have adopted term "quasi-logic gate". Additionally, while $y_{01}$ and $y_{10}$ may both be classified as high based on their magnitude, the two output signals are $\approx \pi$ out-of-phase. This is due to the linearity of the system which imposes the following constraint upon the range of obtainable $y_{out}$ values:

$$y_{out}(x_1 = L_1, x_2 = L_2, x_3, \dots) + y_{out}(x_1 = H_1, x_2 = H_2, x_3, \dots) \qquad (5)$$
$$= y_{out}(x_1 = L_1, x_2 = H_2, x_3, \dots) + y_{out}(x_1 = H_1, x_2 = L_2, x_3, \dots)$$
$$= y_{out}(x_1 = L_1 + H_1, x_2 = L_2 + H_2, x_3, \dots)$$

where $y_{out}(x_1, x_2, x_3, \dots)$ is the value of $y_{out}$ when Eq. 1 is evaluated with the input signals $[x_1, x_2, x_3, \dots]$. To understand the origin of Eq. 5, consider $[x_1, x_2, x_3, \dots]$ to be points within an $N$-dimentional parameter space. The five terms in Eq. 5 are the values of $y_{out}$ evaluated at different points within this space. Since $y_{out}$ is linear, evaluating $y_{out}$ at one point (point 1), then adding the value of $y_{out}$ at another (point 2) is equivalent to moving the evaluation point from point 1 to point 1 + point 2. This is not generally the case for non-linear functions[87]. For instance, consider the two-dimensional non-linear function $y(x_1, x_2) = x_1 x_2^2$. If point 1 is $[x_1 = 1, x_2 = 1]$ and point 2 is $[x'_1 = 1, x'_2 = 2]$ then for point 1 $y(1,1) = 1$ and point 2 $y(1,2) = 4$. But if we now combine point 1 and point 2 we can write $y(x_1 + x'_1, x_2 + x'_2) \neq y(x_1, x_2) + y(x'_1, x'_2)$; i.e., $[y(1 + 1, 1 + 2) = 18] \neq [y(1,1) + y(1,2) = 5]$. The three equations in Eq. 5 represent three different paths in the parameter space which may be taken to the same point $[x_1 = H_1 + L_1, x_2 = H_2 + L_2, x_3, ...]$. Regarding computation, Eq. 5 means that in a linear system the value of $y_{out}$ for each input combination cannot be controlled independently. i.e. one may define the output values of two combinations, for instance $y(x_1 = L_1, x_2 = L_2, x_3, ...) = 0$, $y(x_1 = H_1, x_2 = H_2, x_3, ...) = 4$. However doing so forces the remaining input combinations to be linked with $y(x_1 = L_1, x_2 = H_2, x_3, ...) + y(x_1 = H_1, x_2 = L_2, x_3, ...) = 4$. It is important to note here that $x_3$ to $x_N$ are constants in these functions but have been included to show how this extends to an $N$-input system. This may pose a challenge when cascading operations together because it means that, as it is known, one cannot generally pass an output from one linear logic operation into the input of another while maintaining the functionality of both[78,92]. To do this, an additional non-linear cascading device would be required[92], however this is outside the scope of this work.

To implement the $OR$ operation shown in the right panel of Fig. 2b the following encoding scheme and structural parameters are used: $L_1 = L_2 = 0$, $H_1 = H_2 = 1 \angle 0$ rad (now normalized with respect to $H_1$), $d_1 = 4\lambda_0/3$ and $d_2 = \lambda_0$. It is clear from Eq. 3 that with this encoding scheme $y_{00} = 0$, as no signals are excited at any of the input ports. This is the only output value which should be classified as low, meaning that an $OR$-like operation may be realized if $|y_{01}|, |y_{10}|$ and $|y_{11}| > 0$. Conceptually, this means that any $d_1$ and $d_2$ lengths with any non-zero $H_1$ and $H_2$ encodings may be used to emulate an $OR$ gate so long as the two high signals do not completely destructively interfere at the junction. A theoretical $C_R$ value for this operation is difficult to define as $Th_L = 0$ in Eq. 2. Therefore, instead of selecting $H_1, H_2, d_1$ and $d_2$ to maximize $C_R$, the values listed above are chosen to minimize the range of the high classification region. Using Eq. 3, the theoretical high output states are: $|y_{01}| = |y_{10}| = |y_{11}| = 2/3$. Thus, the theoretical range of the high classification region is 0. If, for instance $L_1 = L_2 = 0$, $H_1 = H_2 = 1 \angle 0$ rad (normalized to $H_1$) and $d_1 = d_2$ were chosen instead, the $OR$ operation can still be realized. However, there would be a wider range of potential output signals. Using this encoding scheme, the theoretical high outputs would be $|y_{01}| = |y_{10}| = 2/3$ and $|y_{11}| = 4/3$, meaning the range of the high classification region would be 2/3. Now, returning to the previous parameter set, where $d_1 =$ $4\lambda_0/3$ and $d_2 = \lambda_0$, numerical simulations of the calculated power distributions and $y_{out}$ values for the four input combinations are presented in the right panels of Fig. 2b. The calculated $y_{out}$ values are: $|y_{00}| = 0$ (no input signals excited), $|y_{01}| = |y_{10}| = 0.666$ and $|y_{11}| = 0.665$, which are in agreement with the theoretical values above.

To implement the *AND* and *NAND* operations shown in Fig. 2d, the parameter sets $L_1 = L_2 = 0$, $H_1 = H_2 = 1 \angle 0$ rad (normalized with respect to $H_1$), $d_1 = d_2 = \lambda_0$ and $L_1 = L_2 = 0$, $H_1 = H_2 = 1 \angle 0$ rad, $d_1 = 4\lambda_0/3$, $d_2 = \lambda_0$ are used, respectively. For both operations, the input signals are normalized with respect to $H_1$. These operations are implemented using the structure presented in Fig. 2c. In this structure there is now a control signal $x_C$ from the top waveguide which has a length $d_C$. These parameters were $x_C = 1/2 \angle 0$ rad, $d_C = 3\lambda_0/2$ and $x_C = 1 \angle 0$ rad, $d_C = 5\lambda_0/3$ for the *AND* and *NAND* operations, respectively. First consider the implementation of the *AND* operation. Using the parameters listed above with Eq. 4 the theoretical values of $y_{out}$ are $|y_{00}| = |y_{01}| = |y_{10}| = 0.25$ and $|y_{11}| = 0.75$. As can be seen the high $y_{11}$ value is clearly distinguishable from the low $y_{00}$, $y_{01}$ and $y_{10}$ values with a theoretical contrast ratio from Eq. 2 of 9.54 dB. Due to the constraints imposed by Eq. 5 this is the maximum theoretical contrast ratio for a linear *AND* gate[77]. Numerical simulations of this scenario are presented in the left panels of Fig. 2d. Here the numerically calculated output signals were $|y_{00}| = 0.25$, $|y_{01}| = |y_{10}| = 0.252$ and $|y_{11}| = 0.751$, meaning the numerical contrast ratio is 9.5 dB, with $Th_H = 0.751$ and $Th_L = 0.252$.

Now consider the *NAND* operation. This implementation is similar to the *OR* gate discussed above however in this scenario the control signal is used to shift the low output signal from the 00 to the 11 input combination. Numerical simulations of this scenario are presented in the righthand panels of Fig. 2d. The calculated numerical output signals for this operation were $|y_{00}| = 0.5$, $|y_{01}| = 0.541$, $|y_{10}| = 0.459$ and $|y_{11}| = 0.0473$, which is close to the theoretical values from Eq. 4 of $|y_{00}| = |y_{01}| = |y_{10}| = 0.5$ and $|y_{11}| = 0$. The calculated contrast ratio of this operation using $Th_L = 0.0473$ and $Th_H = 0.459$ is 19.74 dB.

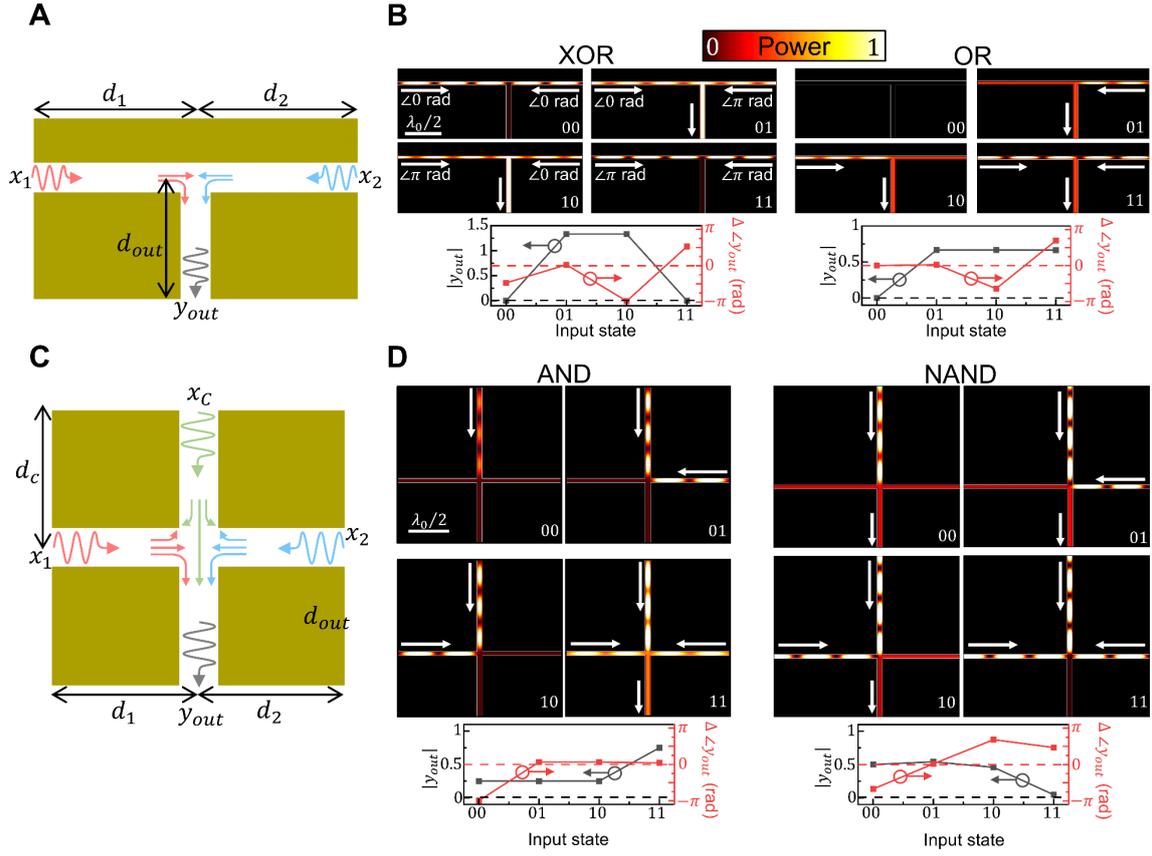

**Fig 2: 2-input linear logic gates. a, c,** Schematic representation of a 2-input 1-output linear logic gate without and with a control signal, respectively. **b, d,** (Top four panels) Normalized power distribution for the 2-input structures when excited with the four possible input combinations (00, 01, 10, 11). (Bottom panel) The output signal (magnitude and phase) calculated across the bottom (output) waveguide for each input combination. **b,** 2-input linear logic gates implemented without a control signal (*XOR* left, *OR* right). **d,** 2-input linear logic gates implemented with a control signal (*AND* left, *NAND* right). Here the phase change at the output $\Delta\angle y_{out}$ is calculated with respect to a $1\angle 0$ rad signal applied from port 1.

**Many-to-one quasi-digital logic gates**

As discussed above, higher order operations with many inputs can be performed by cascading elements, such as logic gates in digital electronics[98]. This is a challenge for quasi-digital structures using analogue signals due to the restrictions imposed by Eq. 5. For instance, a 2-input operation will have at least three unique output signals. These three signals are then classified into two states (high or low), meaning that at least one state will always contain a minimum of two unique signals. When cascading these output states into another quasi-digital linear logic operation, one must consider all potential output signals and design the structure accordingly. This task becomes increasingly challenging as the number of cascaded operations increases, due to the rapid growth in the total number of potential output signals.

One method to overcome this is to consider higher-order operations, not as a system of cascaded elementary operations, but instead as a single quasi-digital operation. The structure, encoding schemes, and control signals can then be optimized to best implement the higher-order operation directly. For instance, consider the implementation of an $N$-input quasi-digital $AND$ gate (an $N$-input operation which should only return a high output signal when all input signals are in the high state). Using Eq. 1, this operation can be realized by designing a structure and selecting an encoding scheme such that $L_a = 0$ and $A_{out,a} H_a = \zeta$, where $\zeta$ is the output signal due to any single high input bit (i.e. each high signal produces the same output signal regardless of the source). Using this encoding scheme, high input signals will constructively interfere with each other, increasing $y_{out}$ by $\zeta$ per high input. This means that $y_{out}$ only depends on the total number of high input signals $N_H$ (irrespective of the permutation) with $|y_{out}| = N_H |\zeta|$. Clearly, $|y_{out}|$ is maximized when all input signals are in the high state ($N_H = N$), allowing for the structure to act as a quasi-digital linear $AND$ gate with $Th_H = N|\zeta|$. Of the remaining output signals (which should all be classified as low), the largest output signal occurs when only one input is in the low state ($N_H = N - 1$). Therefore, the theoretical upper threshold of the low classification region is $Th_L = |\zeta|(N - 1)$ and the theoretical contrast ratio is $C_R = 20 \log_{10}[N/(N - 1)]$. This means that $C_R$ will decrease as the total number of inputs increases. However, like the 2-input case presented in Fig. 2, this can be improved by introducing a control signal, which has been optimized to best implement the $N$-input operation (i.e. different than cascading multiple 2-input quasi-digital $AND$ operations).

This control signal should be carefully chosen so that it destructively interferes with the largest signal which is classified as low (i.e. the signal from which $Th_L$ is defined), thus lowering the value of $Th_L$ and in doing so increasing $C_R$. It should also be noted that $Th_H$ is also reduced due to destructive interference with the control signal. However, since $Th_H > Th_L$ the reduction in $Th_L$ is more significant and the value of $C_R$ increases, which is a desired performance. It is also important to consider the magnitude of the control signal, since when all input bits are low ($N_H = 0$), only the control signal is excited (due to the choice of $L_a = 0$). Thus, the new output signal for this combination is $|y_{out}| = |A_{out,c} x_C|$. If $|x_C|$ is too large, then the magnitude of the all-low output state ($N_H = 0$) may increase above the magnitude of the output state when $N_H = N - 1$. In this case, $Th_L$ should be extracted from the all-low output state (since $Th_L$ is the magnitude of the largest output signal which is classified as low), thus reducing the value of $C_R$ (as $Th_L$ increases). Specifically, for the $N$-input $AND$ operation, $C_R$ is maximized when $x_C$ is chosen such that the magnitude of these two possible output signals ($|y_{out}| = |A_{out,c} x_C|$ when all inputs are low and $|y_{out}| = |\zeta|(N - 1) - |A_{out,c} x_C|$ when $N - 1$ inputs are high, respectively) are equal, as follows:

$$|A_{out,C} x_C| = \frac{|\zeta|(N-1)}{2} \quad (6)$$

The theoretical maximum $C_R$ is for this operation is calculated using Eq. 2 by substituting the new values of $Th_L = |\zeta|(N-1)/2$ and $Th_H = |\zeta|N - |\zeta|(N-2)/2$:

$$C_R = 20 \log_{10}\left(\frac{2N - (N-1)}{N-1}\right) \quad (7)$$

To validate these theoretical calculations, numerical simulations of 3-, 4- and 8-input quasi-digital linear *AND* operations were conducted. The results of these studies are presented in Fig. 3, with the schematic representations of the structures shown in the leftmost panels of Fig. 3a-c, respectively. In each case the structure is made from layers of T-junctions between parallel plate waveguides (and X-junctions where appropriate) combining multiple input waveguides (plus a control waveguide) into a single output waveguide. As before, each junction is connected in a series configuration with all waveguides having the same characteristic impedance $Z_0$ (except for the waveguides which act as the output of a T- or X-junction, which are represented as purple and red waveguides in Fig. 3a-c). These latter waveguides have a characteristic impedance of $\sqrt{2}Z_0$ and $\sqrt{3}Z_0$, respectively and a length of $\lambda_0/4$. They have been included to enable impedance matching between the output waveguides of the T- and X-junctions and the total input impedance of the junction as seen from the output waveguide. With this setup, any backwards traveling signals (towards the input waveguides) produced at the waveguide junctions are directed out of the structure, thereby preventing interference with the quasi-digital logic operation. In practice, these unwanted signals can then be eliminated by adding a circulator to each of the input waveguides, thus preventing them from potentially damaging the sources.

By considering the splitting and superposition of waves at the junctions and accounting for the propagation through the waveguides, $y_{out}$ for the 3-, 4- and 8-input structures is calculated, respectively, as follows:

$$y_{out} = \frac{1}{\sqrt{3}} x_1 e^{i\beta l_1} + \frac{1}{\sqrt{6}}(x_2 e^{i\beta l_2} + x_3 e^{i\beta l_3}) + \frac{1}{\sqrt{3}} x_C e^{i\beta l_C} \quad (8a)$$

$$y_{out} = \frac{1}{\sqrt{8}}(x_1 e^{i\beta l_1} + x_2 e^{i\beta l_2} + x_3 e^{i\beta l_3} + x_4 e^{i\beta l_4}) + \frac{1}{\sqrt{2}} x_C e^{i\beta l_C} \quad (8b)$$

$$y_{out} = \frac{1}{\sqrt{12}}(x_1 e^{i\beta l_1} + x_2 e^{i\beta l_2} + x_3 e^{i\beta l_3} + x_4 e^{i\beta l_4} + x_5 e^{i\beta l_5} + x_6 e^{i\beta l_6} + x_7 e^{i\beta l_7} + x_8 e^{i\beta l_8}) + \frac{1}{\sqrt{3}} x_C e^{i\beta l_C} \quad (8c)$$

where, $l_a$ is the total waveguide length from input $a$ to the output port and $l_C$ is the total waveguide length from the control input to the output port. For the *N*-input quasi-digital *AND* operation $l_1 = l_2 \ldots =$ $l_a = l_c$ is chosen so that the high input signals constructively interfere. From Eq. 8b,c, $(A_{out,a}, A_{out,c}) = (1/\sqrt{8}, 1/\sqrt{2})$ and $(1/\sqrt{12}, 1/\sqrt{3})$ for the 4- and 8-input structures, respectively. This means that the simple encoding scheme $L_a = 0, H_a = 1 \angle 0$ rad can be used to implement and *AND* operation with $A_{out,a} = \zeta = 1/\sqrt{8}$ and $1/\sqrt{12}$ for the 4- and 8-input structures, respectively (i.e., since the scattering towards the output is the same from each input, the same encoding scheme may be used for each input bit). For the 3-input structure $A_{out,1} = 1/\sqrt{3}$, $A_{out,2} = A_{out,3} = 1/\sqrt{6}$ and $A_{out,C} = 1/\sqrt{3}$ (i.e., $A_{out,a}$ is not the same for each $a$). Therefore, to enforce $A_{out,a} H_a = \zeta$, input 1 requires a different encoding scheme than inputs 2 and 3. If $H_2 = H_3 = 1 \angle 0$ rad is chosen then $A_{out,2} H_3 = A_{out,3} H_3 = \zeta = 1/\sqrt{6}$. Therefore, $H_1 = \zeta/A_{out,1} = 1/\sqrt{2}$ is chosen as the encoding scheme for input 1. Using Eq. 6 and selecting input phase to ensure destructive interference with $\zeta$, the calculated optimal values of $x_C$ are $x_C = 1/\sqrt{2} \angle \pi$ rad, $x_C = 3/4 \angle \pi$ rad and $x_C = 7/4 \angle \pi$ rad, for the 3-, 4- and 8-input structures respectively. In all three examples, the input signals are normalized with respect to the $H_3$ input signal.

The numerically calculated power distributions of the 3-, 4-, and 8-input structures using these parameters are presented in the right-hand panels of Fig. 3a-c, respectively. These examples have been selected to represent results for input combinations ranging from all-low to all-high. The corresponding values of $|y_{out}|$ for each structure and input combination are presented in Fig. 3d. It is worth noting that due to the choice of the encoding scheme (chosen such that $A_{out,a} H_a = \zeta$ is constant) $|y_{out}|$ only depends on the total number of high inputs (irrespective of the permutation), meaning that every obtainable value of $|y_{out}|$ is shown in Fig. 3d. In each case, the high output signal (when all inputs are high) is clearly distinguishable from all other output signals. The numerical classification thresholds extracted from Fig. 3d are $Th_H = 0.809, 0.941, 1.480$ and $Th_L = 0.407, 0.588, 1.224$ for the 3-, 4-, and 8-input structures, respectively. These threshold values are marked by red ($Th_H$) and blue ($Th_L$) dashed lines in Fig. 3d. Using Eq. 2, the $C_R$ values of these operations are calculated to be 5.97 dB, 4.08 dB, and 1.67 dB, respectively, which is close to the theoretical values of 6.02 dB, 4.44 dB, and 2.18 dB, calculated using Eq. 7. For completeness, the theoretical and numerical results showing the ratio of $Th_H$ over $Th_L$ for several of quasi-digital $N$-input *AND* operations are presented in Fig. 3e. Here, numerical results are provided for $N = 2$ to $N = 8$. The results corresponding to the $N = 2$ operation are the same as those presented in Fig. 2. Information regarding the structures used to implement the 5-, 6-, and 7-input operations can be found in the supplementary materials. As expected, the $Th_H/Th_L$ is asymptotic to 1 from above ($C_R$ is asymptotic to 0 dB from above) when $N \to \infty$. This means that a quasi-digital linear *AND* operation can be theoretically realized for any number of inputs. However, in practice the maximum number of bits would be limited by noise due to the low value of $C_R$ for large numbers of inputs. This can be seen in Eq. 7 as when $N \to \infty$, $C_R \to 0$.

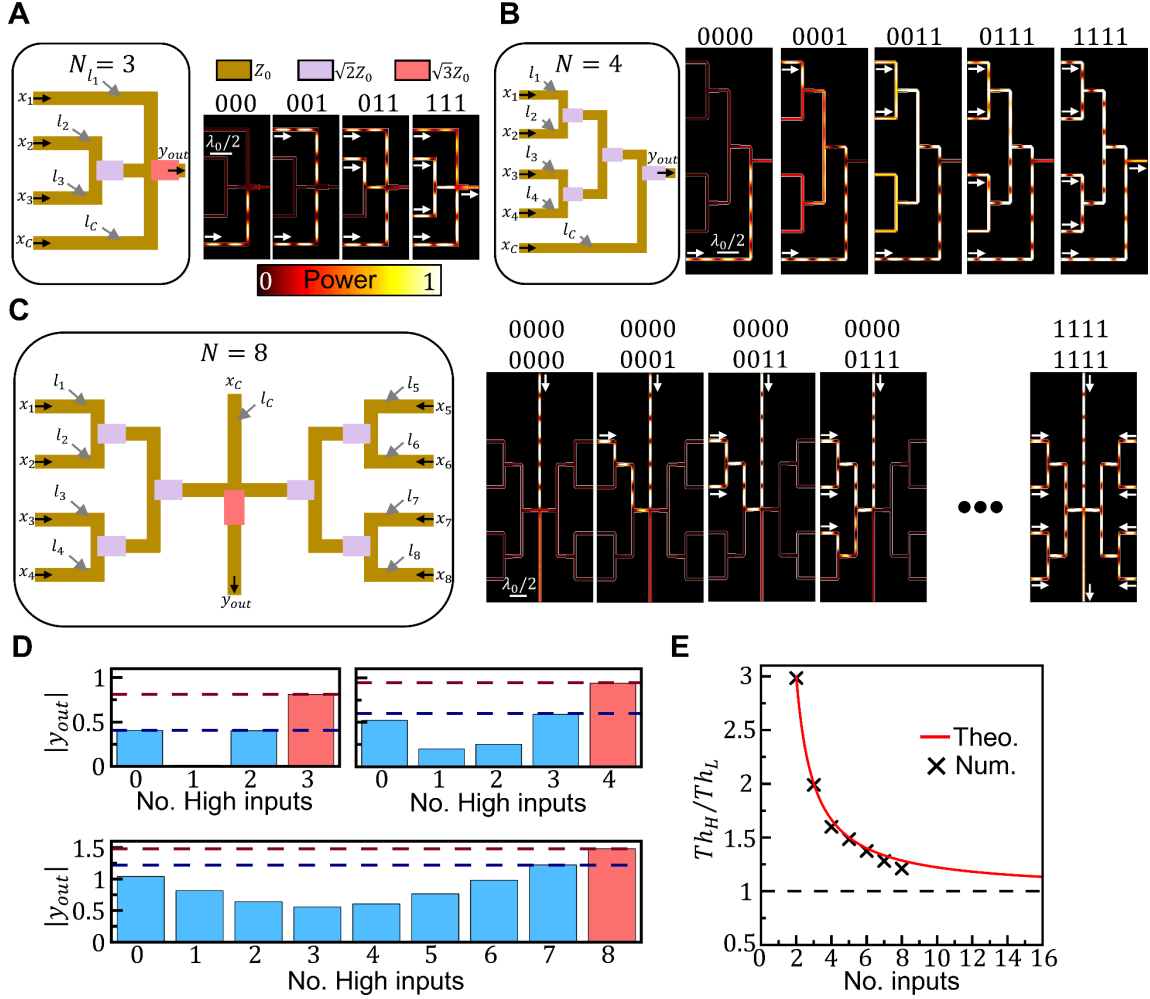

**Fig 3: Many-to-one linear logic gate example: *N*-input AND gate. a**, **b**, **c**, (Left) Schematic representations of the waveguide structure for the $N = 3$, 4 and 8 linear *AND* gates, respectively. (Right) Normalized power distributions of the scenarios on the left for input combinations ranging from all-zero to all-one. Here, waveguides are color coordinated to indicate their characteristic impedance with purple and red waveguides representing working as QWTs connected to the output of a T- and X-junction, respectively. These waveguides have an impedance of $\sqrt{2}Z_0$ and $\sqrt{3}Z_0$ which is achieved by changing the waveguide plate separation from 1 mm to 1.41 mm and 1.73 mm, respectively. **d**, Output signals ($y_{out}$) calculated across the output waveguides as a function of the number of high inputs for the $N = 3$ (top-left), $N = 4$ (top-right) and $N = 8$ (bottom) structures. **e**, Theoretical (red line) and numerical (black crosses) high-low output threshold contrast ratios for *N*-input linear *AND* gates.

**Many-to-many quasi-digital logic gates**

In addition to the many-to-one operations discussed in the previous section, it is also possible to emulate certain many-to-many operations using quasi-digital linear systems. This mean that the structure can apply multiple operations to enable different output signals. In this case, the classification thresholds (and associated $C_R$ values) can then be evaluated separately for each individual operation (an output). This poses a challenge for quasi-digital linear systems as the selected encoding scheme of the input signals must be applicable to the operation required at each output. Additionally, potential multipathing[99,100] must be considered. Specifically, a wave traveling from an input to an output may arrive via a direct or indirect route. These indirect paths may be caused by, for instance, reflections at waveguide junctions. Additionally, when considering multiple signals being applied by different ports, a signal may not be fully directed towards the desired output port, instead being partially transmitted towards another input waveguide. This can be prevented by, for instance, using a quarter-wave-transformer (QWT) as a fan-out operator (wave-splitter) and a Wilkinson power divider[96] (WPD) as a fan-in operator (wave-combiner), as shown in the left and right panels of Fig. 4a, respectively. Importantly, the fan-in operator (WPD) is also designed to combine two input signals while isolating the two input ports from each other. This is done by directing a portion of the input signals towards a pair of *Drop* waveguides where they may be eliminated via properly matched loads[96] (in this case waveguide ports). A schematic representation of the signal paths through the WPD can be seen in the right panel of Fig. 4a.

One example of a many-to-many operation which may be emulated using quasi-digital linear systems is a half-adder[101]. This is a two-input two-output operation where one output (called $SUM$) is the result of an $XOR$ operation and the other (called $CARRY$) is due to an $AND$ operation. A schematic representation of this operation using the considerations explained above is shown in Fig. 4b. Here a pair of fan-out (QWTs) operators split the input signals ($x_1$ and $x_2$) into two separate paths. Signals from either input are then combined via a fan-in (WPD) operation at the $SUM$ and $CARRY$ outputs. Following this, different operations can be realized by controlling the length of the waveguides between the fan-out and fan-in stages. In this case, the length of the waveguides from either fan-out to the $CARRY$ output are the same ($L_1$). Conversely, there is a $\lambda_0/2$ difference in the length of the waveguides towards the $SUM$ output. Due to this, when two input signals constructively interfere at the $CARRY$ output, they instead destructively interfere at the $SUM$ output, and vice versa. Considering the splitting and superposition of waves at the fan-out and fan-in operators and accounting for wave propagation within the waveguides, the *SUM* and *CARRY* output signals ($y_{SUM}$ and $y_{CARRY}$, respectively) are calculated as follows:

$$y_{SUM} = \frac{1}{2}e^{i\beta l_{S,1}}x_1 + \frac{1}{2}e^{i\beta l_{S,2}}x_2 \tag{9a}$$

$$y_{CARRY} = \frac{1}{2}e^{i\beta l_{C,1}}x_1 + \frac{1}{2}e^{i\beta l_{C,2}}x_2 \tag{9b}$$

where $l_{S,1} = l_{C,1} = l_{C,2} = L_1 + \lambda_0/2$ and $l_{S,2} = L_1 + \lambda_0$. Numerical simulation results for power distributions for the case when an incident signal is applied at either input of the quasi-digital half adder network ($x_1$, top and $x_2$, bottom) are presented in Fig. 4c. From these results, it can be observed that indirect routes between the inputs and outputs are eliminated. This is evidenced by the absence of standing waves, indicating no reflections at the junctions as well as the isolation between the left and right halves of the network.

Using Eq. 9a,b, the *SUM* and *CARRY* operations are implemented simultaneously using the input encoding scheme $L_1 = L_2 = 1/3 \angle \pi$ rad and $H_1 = H_2 = 1 \angle 0$ rad. Similar to the elementary *XOR* presented in Fig. 2b, the *SUM* operation is realized due to the complete destructive interference of like (representing the same input state) input signals. The *CARRY* operation is realized due to the constructive interference of the two high input signals. The magnitude ratio of the high and low input signals is then selected to maximize the $C_R$ value of the *CARRY* operation (the *SUM* operation is already maximized due to the choice of phase). Numerical simulation results of the power distributions and the output signals of this network are presented in Fig. 4d,e, respectively. As observed, the *SUM* and *CARRY* operations are successfully emulated with the numerical threshold values $Th_L = 0.0461$, $Th_H = 0.655$ and $Th_L = 0.366$, $Th_H = 1.066$, respectively. Using Eq. 2, this corresponds to $C_R$ values of 23.45 dB and 9.2877 dB, for the *SUM* and *CARRY* operations, respectively.

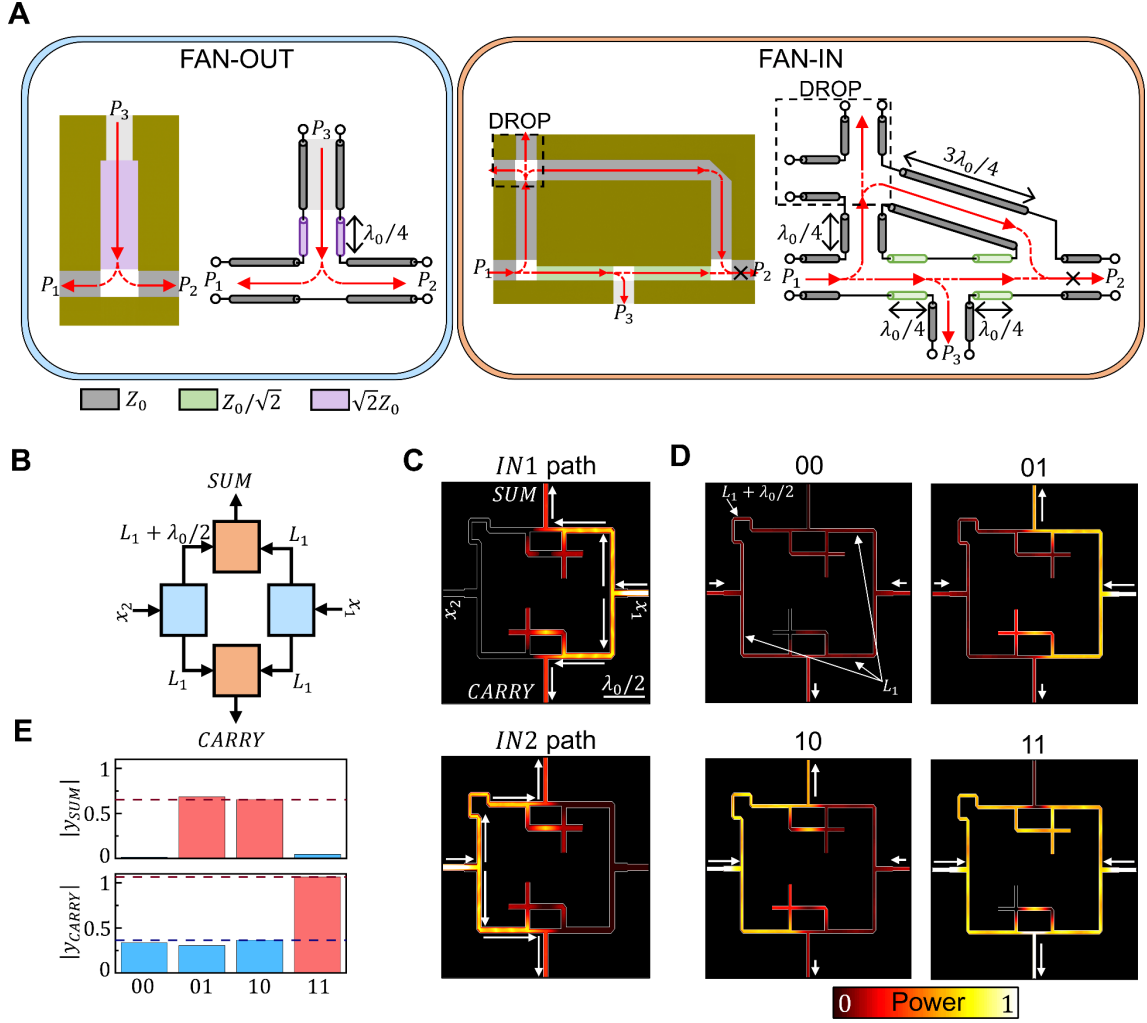

**Fig 4: EM wave-based linear half-adder. a**, Schematic representation of linear fan-out (left) and fan-in (right) operations made from a quarter-wave-transformer and a Wilkinson power divider, respectively. The waveguide structure and the TL schematic are shown in the left and right subpanels, respectively. **b**, Schematic representation of an EM wave-based half-adder. Here the blue and red squared represent a fan-out and fan-in operation, respectively. **c**, Normalized power distribution showing the signal path routes through the linear half-adder from input 1 (top) and input 2 (bottom), respectively. **d**, Power distribution for the four input combinations (00, 01, 10, 11). **e**, Output signal observed at the $SUM$ (top) and $CARRY$ (bottom) waveguides, respectively.

To further demonstrate the potential of many-to-many quasi-digital linear logic operations, the half-adder structure presented in Fig. 4 is modified to emulate a full-adder and a two-bit full-adder. Schematic representations of the modified structures are presented in Fig. 5a and Fig. 6a, respectively. To implement the full-adder (Fig. 5a), the half-adder structure is altered to include a third input. Signals from this input are split via a fan-out operator (QWT) and combined with the outputs from the half-adder via a fan-in operator (WPD) to produce a new pair of $SUM$ and $CARRY$ outputs. Considering the fan-in operators, the fan-out operators and the signal routes through the network, the $SUM$ and $CARRY$ output signals can be calculated as:

$$y_{SUM} = \frac{1}{2\sqrt{2}} e^{i\beta l_{S,1}} x_1 - \frac{1}{2\sqrt{2}} e^{i\beta l_{S,2}} x_2 + \frac{1}{2} e^{i\beta l_{S,3}} x_3 \qquad (10a)$$

$$y_{CARRY} = \frac{1}{2\sqrt{2}} e^{i\beta l_{C,1}} x_1 + \frac{1}{2\sqrt{2}} e^{i\beta l_{C,2}} x_2 + \frac{1}{2} e^{i\beta l_{C,3}} x_3 \qquad (10b)$$

where $l_{S,1} = l_{C,1} = l_{C,2} = l_{S,3} = l_{S,4} = 2L_1 + 3\lambda_0/4$ and $l_{S,2} = 2L_1 + 5\lambda_0/4$. The operating principle of the full-adder requires that, when the third input (input 3) is in the low state ($x_3 = L_3$) the $SUM$ and $CARRY$ outputs should resemble an $XOR$ and $AND$-like operation, respectively, between inputs 1 and 2. This can be achieved by selecting $L_3 = 0$ and using the same encoding scheme as the half-adder for inputs 1 and 2. However, when input 3 is in the high state ($x_3 = H_3$) the response should shift to instead emulate an $XNOR$ and $OR$-like operation at the $SUM$ and $CARRY$ outputs, respectively. Using Eq. 10a,b this can be implemented by choosing $H_3 = 1/\sqrt{2} \angle 0$ rad for input 3. An in-depth explanation for this choice of $H_3$ value can be found in the supplementary materials. Numerical simulation results showing the output signals produced by this structure are presented in Fig. 5b. As expected, the output signals obtained when $x_3 = L_3 = 0$ closely resemble the half-adder results presented in Fig. 4e. When $x_3 = H_3 = 1/\sqrt{2} \angle 0$ rad, the shift from $AND$-like to $OR$-like behavior for the $CARRY$ output is clearly visible. For the $SUM$ output, the signal corresponding to the 10 input combination between inputs 1 and 2 (ordered $x_2, x_1$) is reduced into the low classification region, while the signals from the 00 and 11 input combinations appear into the high classification region, as expected. However, the $SUM$ output signal from the 01 input combination has instead been raised above the high classification region. This is a consequence of the linearity of the system, meaning that it is impossible to define a value of $H_3$ which destructively interferes with the output signals from both 01 and 10 input combinations. The $XNOR$-like performance is still emulated; however, this requires the implementation of a second low classification region (with its own associated threshold value) positioned above the high classification region. Consequently, it is also necessary to define an upper threshold for the high classification region. The $C_R$ value of this operation is found by evaluating Eq. 2 for each combination of high and low threshold values and selecting the smallest $C_R$ value (considering the magnitude of $C_R$ if $Th_L > Th_H$). Numerical simulation results for the power distributions when $x_3 = H_3$ are presented in Fig. 5c. For completeness the numerical results when $x_3 = L_3$ are presented in the supplementary materials. The numerically calculated threshold values of the $CARRY$ output are $Th_L = 0.266$ and $Th_H = 0.7$, corresponding to $C_R = 8.4$ dB. This value is close to, but slightly lower than, that of the half-adder's $CARRY$ operation from Fig. 4e ($C_R = 9.2877$ dB). For the $SUM$ operation, the threshold values for the upper and lower low classification regions are $Th_L = 0.0251$ and $Th_L = 0.969$, respectively. The upper and lower threshold values for the high classification region are $Th_H = 0.464$ and $Th_H = 0.498$, respectively. Thus, the $C_R$ value of this operation is calculated using Eq. 2 with $Th_H = 0.498$ and $Th_L = 0.969$, yeilding $C_R = 5.78$ dB. This is an example of a quasi-digital operation where $Th_L > Th_H$. This value is lower than the $C_R$ of the $SUM$ operation for the half-adder from Fig. 4e ($C_R = 23.45$ dB) due to the requirement of a second low classification region.

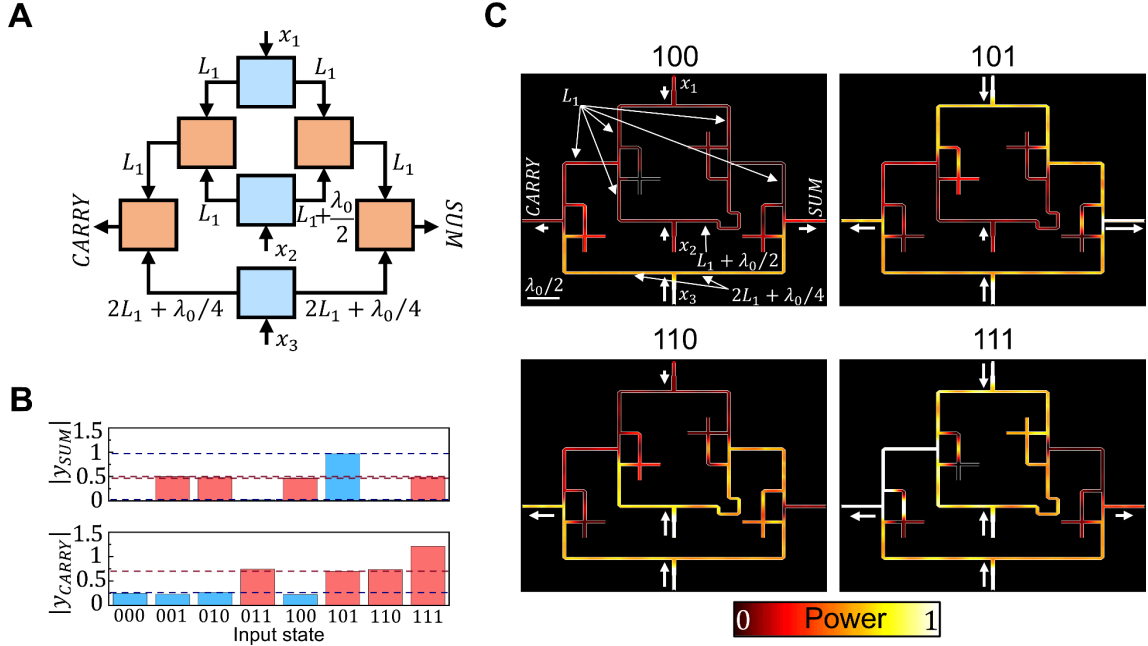

**Fig 5: EM wave-based linear full-adder. a,** Schematic representation of the full-adder constructed from fan-out and fan-in operations. **b,** Output signal calculated at the $SUM$ (top) and $CARRY$ (bottom) waveguides for each input combination. **c,** Normalized power distribution for input combinations between 100 and 111 (ordered $x_3, x_2, x_1$).

A similar structure as the one discussed in Fig. 5 is used to implement the quasi-digital two-bit full-adder. A full-adder is a 4-input, 3-output structure which computes the bit-wise addition of two 2-bit numbers $A$ and $B$[102]. A schematic representation of the structure used to implement this operation is presented in Fig. 6a. Note that the ordering of inputs in Fig. 6a has been chosen to be consistent with the structures presented in Fig. 4a and Fig. 5a. Using this ordering, the individual bits of $A$ and $B$ ($A = [A1, A0]$ and $B = [B1, B0]$) are defined using the following inputs: $A1 \to x_1, B1 \to x_2, A0 \to x_3, B0 \to x_4$. The outputs ($SUM1, SUM2$ and $CARRY$) correspond to the first, second and third bits of the operation $A + B$, respectively. This operation is realized by connecting a quasi-digital half- and a full-adder. Here, the $CARRY$ output of the half-adder is connected to input 3 of the full-adder. The signals seen at each output of the overall two-bit full-adder network are then calculated by considering the splitting and superposition of waves at the fan-out and fan-in operators and accounting for the various signal routes through the network. Full details of this calculation can be found in the supplementary materials.

$$y_{SUM1} = \frac{1}{2} e^{i\beta l_{S1,3}} x_3 + \frac{1}{2} e^{i\beta l_{S1,4}} x_4 \quad (11a)$$

$$y_{SUM2} = \frac{1}{2\sqrt{2}} e^{i\beta l_{S2,1}} x_1 + \frac{1}{2\sqrt{2}} e^{i\beta l_{S2,2}} x_2 + \frac{1}{4} e^{i\beta l_{S2,3}} x_3 + \frac{1}{4} e^{i\beta l_{S2,4}} x_4 \quad (11b)$$

$$y_{CARRY} = \frac{1}{2\sqrt{2}} e^{i\beta l_{C,1}} x_1 + \frac{1}{2\sqrt{2}} e^{i\beta l_{C,2}} x_2 + \frac{1}{4} e^{i\beta l_{C,3}} x_3 + \frac{1}{4} e^{i\beta l_{C,4}} x_4 \quad (11c)$$

where $l_{S1,3} = L_1 + \lambda_0/2$, $l_{S1,3} = L_1 + \lambda_0$, $l_{S2,1} = l_{C,1} = l_{C,2} = 2L_1 + 3\lambda_0/4$ and $l_{S2,2} = l_{C,3} = l_{S2,3} = l_{C,4} = l_{S2,4} = 2L_1 + 5\lambda_0/4$. Here, $SUM1$ (Eq. 11a) is the same as the $SUM$ output from a half-adder (Eq. 9a). Moreover, $SUM2$ (Eq. 11b) and $CARRY$ (Eq. 11c) resemble the $SUM$ (Eq. 10a) and $CARRY$ (Eq. 10b) from a full-adder. However, the terms involving input 3 within Eq. 10a,b have now been replaced with the output of a quasi-digital $AND$ operation between inputs 3 and 4. This is due to the connection between the $CARRY$ output of the half-adder and input 3 of the full-adder. Using Eq. 11 the following encoding scheme is used to implement the 2-bit full-adder operation: $L_1 = L_2 = 1/3 \angle \pi$ rad, $H_1 = H_2 = 1 \angle 0$ rad, $L_3 = L_4 = 1/(3\sqrt{2}) \angle \pi$ and $H_3 = H_4 = 1/\sqrt{2} \angle 0$ rad. An in-depth explanation regarding this choice is presented in the supplementary materials. Full-wave numerical simulation results for the power distribution and output signals of the full-adder structure are presented in Fig. 6b,c, respectively. The power distributions of 6 different input combinations are shown in Fig. 6b as examples. For completeness, simulation results for the remaining 10 input combinations are presented in the supplementary materials. As can be seen in Fig. 6c, the high and low classification regions at all 3 outputs are clearly distinguishable meaning the structure indeed emulates the performance of the 2-bit full-adder. Like the $SUM$ output from the full-adder structure (shown in Fig. 5b), the $SUM2$ output from Fig. 6c also has a single low classified output signal with a larger magnitude than all the high classified output signals. Therefore, for this output it is also necessary to define an upper threshold value for the high classification regions and a second low threshold value. The numerically calculated threshold values for the $SUM1$ and $CARRY$ outputs are $Th_L = 0.0892$, $Th_H = 0.475$ and $Th_L = 0.381$, $Th_H = 0.608$, respectively, corresponding to $C_R$ values of $C_R = 14.54$ dB and $4.07$ dB, respectively. For the $SUM2$ output the low threshold values are $Th_L = 0.183$, $Th_L = 0.839$ and the high threshold values are $Th_H = 0.349$, $Th_H = 0.607$. Thus $C_R$ is calculated using $Th_H = 0.607$ and $Th_L = 0.839$, yielding $C_R = 2.82$ dB. These results demonstrate the potential of linear system to perform certain many-to-many quasi-digital operations. Similar to the many-to-one example provided in Fig. 3, there is a trade-off between the number of inputs and the maximum attainable $C_R$ value of the operation, with an increase in the number of inputs resulting in a reduced maximum $C_R$ value. This trade-off is clearly demonstrated by the results presented in Fig. 4e, 5b and 6c. Additionally, it has also been shown how in some cases, such as the *SUM* output of the full-adder (see Fig. 5b), due to the linearity of the system, it is necessary to define multiple output classification regions which also impacts the $C_R$ value of the operations. Indeed, for operations with more inputs, such as a 4-bit or 8-bit full-adder, it may also be necessary to also define multiple high classification regions.

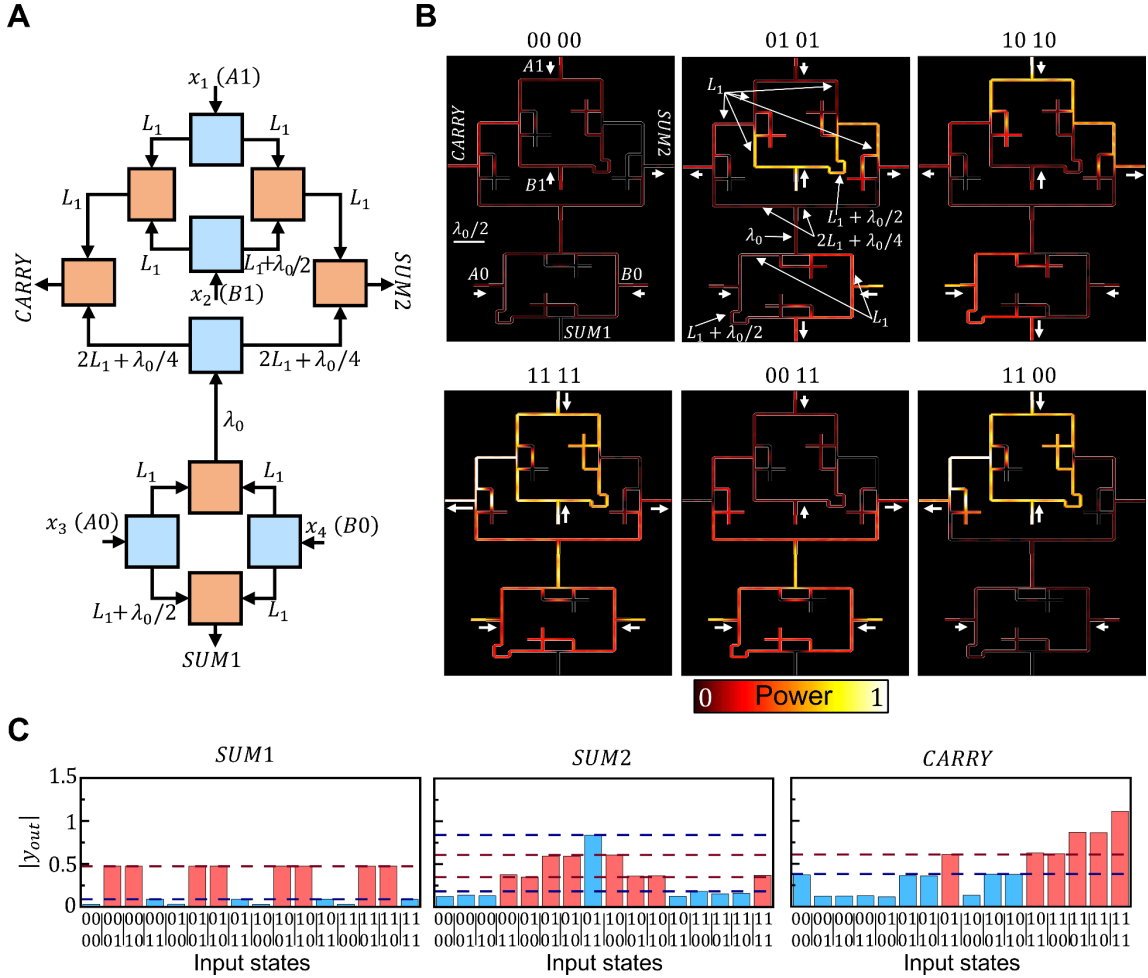

**Fig 6: EM wave-based linear 2-bit adder. a**, Schematic of the 2-bit adder using fan-out and fan-in operations. **b**, Normalized power distribution of the two-bit adder when excited using 6 different input combinations, as examples. The results are normalized with respect to the maximum power distribution when $x_1 = H_1$ is excited at input 1. **c**, Output signal calculated across the *SUM*1 (left), *SUM*2 (middle) and *CARRY* (right) waveguides for all input combinations. The input bits are arranged in the flowing order $A1, B1, A0, B0$ ($x_1, x_2, x_3, x_4$).

## Conclusions

To conclude, this work has explored the potential for quasi-digital computing with EM-waves based on the superposition of signals within linear structures. By carefully designing the encoding scheme, mapping the high or low states of input bits onto the wave characteristics of monochromatic input signals, one may emulate the performance of digital logic operations in a quasi-digital system. Various quasi-digital linear logic operations have been presented throughout this work. This was done by exploiting networks of interconnected parallel plate waveguides at microwave frequencies. The example operations include, elementary 2-input 1-output operations, many-to-one operations and many-to-many operations. In each case, the performance of these operations has been verified via full-wave numerical simulations, which are in agreement with the theoretical calculations. We envision that the work presented in this manuscript could enable new opportunities for high-speed quasi-digital EM wave-matter computing systems.

## Methods

The numerical results presented throughout this work have been obtained using the frequency domain solver of the commercial full-wave simulation software CST Studio Suite®. In each case, the waveguide networks were constructed using waveguides with a $1 \times 1$ mm cross-section, connected at junctions in the series configuration. The metallic plates were modeled as a thin layer (0 thickness) of PEC. Vacuum ($\varepsilon_r = 1$, $\mu_r = 1$) was used as the waveguide filling material. Where necessary, changes in waveguide impedance has been realized by varying the separation between the metallic plates. The various input and output signals used throughout this work are excited via waveguide ports placed at the ends of the input and output waveguides, respectively. The boundary conditions used in these simulations are *open* for positive and negative $x$ and $y$ (top, right, bottom and left when viewed from above as in Figs. 2-6) and *open (add space)* for positive and negative $z$ (into- and out-of-plane as seen in Figs. 2-6), respectively.

## Acknowledgements

This work was supported by the Leverhulme Trust under the Leverhulme Trust Research Project Grant scheme (RPG-2020-316, RPG-2023-024).. For the purpose of Open Access, the authors have applied a CC BY public copyright license to any Author Accepted Manuscript (AAM) version arising from this submission.

## Conflicts of interests

The authors declare no conflicts of interests.

## Data availability

The datasets generated and analyzed during the current study are available from the corresponding author.